\documentclass[preprint,12pt,3p]{elsarticle}

\usepackage[T1]{fontenc}
\usepackage[english]{babel}
\usepackage{hyperref}
\usepackage{ifpdf}
\usepackage{subfigure}
\usepackage{amssymb}
\usepackage{amsfonts}
\usepackage{epsf}
\usepackage{rotating}
\usepackage{graphicx}
\usepackage{amsmath}
\usepackage{fancyhdr}
\usepackage{lineno}
\usepackage{dutchcal}
\usepackage{babel}
\usepackage{graphics}
\usepackage{pstricks}
\usepackage{color}
\usepackage{multirow}

\newcommand{\nn}{\nonumber}
\newcommand{\lsim}{\mathrel{\mathop{\kern 0pt \rlap
  {\raise.2ex\hbox{$<$}}}
  \lower.9ex\hbox{\kern-.190em $\sim$}}}
\newcommand{\gsim}{\mathrel{\mathop{\kern 0pt \rlap
  {\raise.2ex\hbox{$>$}}}
  \lower.9ex\hbox{\kern-.190em $\sim$}}}

\newcommand{\be}{\begin{equation}}
\newcommand{\ee}{\end{equation}}
\newcommand{\bea}{\begin{eqnarray}}
\newcommand{\eea}{\end{eqnarray}}

\begin{document}

\begin{frontmatter}

\title{\boldmath Bilepton Signatures at the LHC}
%\tnotetext[label0]{This is only an example}

\author[label1]{Gennaro Corcella}
\address[label1]{INFN, Laboratori Nazionali di Frascati,
  \\Via E.~Fermi 40, 00044, Frascati (RM), Italy}
\ead{gennaro.corcella@lnf.infn.it}

\author[label5]{Claudio Corian\`o}
\address[label5]{Dipartimento di Matematica e Fisica "Ennio De Giorgi", Universit\`a del Salento and INFN-Lecce, \\ Via Arnesano, 73100 Lecce, Italy}
\ead{claudio.coriano@le.infn.it}

\author[label5]{Antonio Costantini}
%\address[label5]{Dipartimento di Matematica e Fisica "Ennio De Giorgi", Universit\`a del Salento and INFN-Lecce, \\ Via Arnesano, 73100 Lecce, Italy}
\ead{antonio.costantini@le.infn.it}

\author[label5]{Paul H. Frampton}
%\address[label5]{Dipartimento di Matematica e Fisica "Ennio De Giorgi", Universit\`a del Salento and INFN-Lecce, \\ Via Arnesano, 73100 Lecce, Italy}
\ead{paul.h.frampton@gmail.com}

\begin{abstract}
We discuss the main signatures of the Bilepton Model at the Large Hadron Collider, focusing on its gauge boson sector. The model is characterised by five additional gauge bosons, 
  four charged and one neutral, beyond those of the Standard Model, plus
  three exotic quarks. The latter turn into ordinary quarks with the emission of bilepton doublets $(Y^{++},Y^{+})$ and $(Y^{--},Y^{-})$ of lepton number  $L=-2$ and $L=+2$ respectively, with the doubly-charged bileptons decaying into
  same-sign lepton pairs.
  We perform a phenomenological analysis investigating
  processes with two doubly-charged bileptons and
  two jets at the LHC and find that, setting suitable cuts on pseudorapidities
  and transverse momenta of final-states jets and leptons,
  the model yields a visible signal and 
  the main Standard Model backgrounds can be suppressed.
  Compared to previous studies, our investigation is based on a  full Monte
  Carlo implementation of the model and accounts for parton showers,
  hadronization and
  an actual jet-clustering algorithm for both signal and 
  Standard Model background, thus providing an optimal framework for
   an actual experimental search.
  
  \end{abstract}

\end{frontmatter}

%\flushbottom

\section{Introduction}

The discovery of the Brout-Englert-Higgs (BEH) boson
\cite{HiggsCMS,HiggsATLAS} 
has marked a major advance in high-energy physics and neatly completes \cite{Higgs1, Higgs2,BE} the particle
content of the Standard Model (SM) in its current formulation. Through 
the decades before the confirmation of the BEH (Higgs) mechanism, a wide
variety of Beyond the Standard Model (BSM) theories 
have been advanced \cite{horava}, usually
involving new particles not contained in the SM. The most frequently
cited reason for proceeding
to BSM from the SM has been the issue of the naturalness of the Higgs mass 
when quadratic divergences suggest its being much heavier than observed.

The most popular BSM proposal \cite{WZ1,WZ2} has been, for decades,
weak-scale supersymmetry; though not being ruled out yet, the
LHC data now available regretfully offer no encouragement to supersymmetry.
A second popular scenario is based on large extra dimensions,
most forcefully extolled in \cite{ADD}, but also with a negative
outcome according to current data.

If we abandon the directions of weak scale supersymmetry and large extra dimensions, a more conservative assumption
is that the most appropriate BSM theory is a renormalizable gauge field model 
conceptually identical
to the SM but with extra states. One expects the BSM to be motivated by
some facts unexplained in the SM and to be testable by experiment. One 
striking feature of the SM is the existence of three quark-lepton families when the first
family alone accounts for the vast majority of the baryonic material while
the second and third generations appear at first sight as a curious redundancy.
Nevertheless, the family replication is crying out to be a clue.
There is an infinite number of possibilities for such a model, selected
by the choice of gauge groups and
irrreducible representations of the chiral fermions,
but the requirement
of motivation and testability reduce this to a small finite choice.

The simplest of
such models, to our knowledge, is the Bilepton Model\footnote{We change
name from 331-model because it is necessary to specify not only the gauge group
but choices of matter representations and electric charge embedding. We change
nomenclature only to avoid confusion.} of \cite{PHF,PP}
where the occurrence of three families is underwritten by cancellation of
triangle anomalies\footnote{An important and prescient precursor of the
  Bilepton
Model was made in 
1980 \cite{Valle} in a model where, however, the embedding of electric charge
does not accommodate doubly-charged bileptonic gauge bosons.}. One family, the most massive, is 
treated asymmetrically\footnote{The difference between the otherwise identical models
  of \cite{PHF} and \cite{PP} is that in the latter
  it is the first fermion family which is
treated asymmetrically, not the third, a
choice which does not allow adequate suppression of flavor-changing neutral currents.}
and although each family has a non-zero anomaly, the three families
combine together to give a vanishing anomaly as essential for consistency.

\section{The Bilepton Model}

When the Bilepton Model was introduced,
it seemed possible that it was 
merely one of a large class of such model and therefore had a small
probability of being correct.  Other than 
redefinitions of the charge operator, no alternative has been discovered
in the intervening 25 years, and therefore
it appears more unique than originally
thought and thus much more worthy of serious consideration. Here we shall
stay with the original formulation \cite{PHF} with its charge operator
which includes {\em bileptons}, coupled to same-sign leptons.

The model introduces three types of new particles beyond the SM:
gauge bosons, exotic quarks and additional scalars. There are five
extra gauge bosons which include one $Z'$ and four bileptons in two
$SU(2)_L$ doublets $(Y^{--}, Y^{-})$, with lepton number $L=+2$,
as well as $(Y^{++},Y^{+})$
with $L=-2$.
Each bilepton will decay to two same sign leptons,
while exhibiting a coupling  to one SM quark and to an exotic quark
of the same family, 
the latter denoted as $D, S,$ and $T$.
The consistency
of the model is quite remarkable \cite{meirose1}:
we can anticipate that, for the processes
which will be investigated in the following, leading to the
production of two same-sign lepton pairs and two or more jets, 
there are about 3000 amplitudes contributing, and the analysis
therefore has been automatized. We have implemented the model
into \texttt{SARAH 4.9.3} \cite{sarah}, with the amplitudes
computed numerically using \texttt{MadGraph} \cite{madgraph}.
The simulation of parton showers and hadronization 
has been carried out by using \texttt{HERWIG} \cite{herwig}.  

In this article, our principal focus is on the pair production of these doubly-charged
bileptons $Y^{--}Y^{++}$ and their subsequent decay into like-sign lepton
pairs, for which we present a detailed simulation of the corresponding events and 
SM background. As in previous analysis, the $Z'$ of the bilepton model is leptophobic \cite{Dumm} and 
possesses a decay width comparable to its mass, rendering it an
ill-defined resonance with respect to empirical verification. As already mentioned above, four major variants of the model have been discussed recently from the LHC perspective, identified by the value of the charge operator 
$Q=T^3 +\bar{ \beta} T^8 + X$, with $(\bar{\beta}=\pm \sqrt{3},\pm1/\sqrt{3})$ \cite{cao}. We will consider the original model of \cite{PHF} with $\bar{\beta}=\sqrt{3}$ which allows doubly-charged bileptons in the spectrum.
Notice that only the choices $\pm \sqrt{3}$ allow doubly-charged bileptons. We do not
discuss the production and decays of new scalars because those results
will be less specific to the model considered in this paper.

\section{Theoretical Framework}

As already stated above, the gauge structure of the bilepton model of
\cite{PHF,PP} is $SU(3)_c \times SU(3)_L \times 
U(1)_X$, with the fermions in the fundamental of $SU(3)_c$ arranged into triplets of $SU(3)_L$. As already pointed out, the three families of quarks are treated asymmetrically with respect to the weak $SU(3)$ $(SU(3)_L)$, with the first two families given by
\begin{equation}
Q_1=\left(
\begin{array}{c}
u_L\\
d_L\\
D_L
\end{array}
\right),\quad Q_2=\left(
\begin{array}{c}
c_L\\
s_L\\
S_L
\end{array}
\right),\quad Q_{1,2}\in(3, 3, -1/3)
\end{equation}
under $SU(3)_c \times SU(3)_L \times 
U(1)_X$,
whereas the third family is
\begin{equation}
Q_3=\left(
\begin{array}{c}
b_L\\
t_L\\
T_L
\end{array}
\right),\quad Q_3\in(3,\bar3, 2/3).
\end{equation}
$D, S$ and $T$ are exotic (extra) quarks, which in our simulation will be in the TeV (1.1-1.3 TeV) mass range. 
The right-handed quarks ($\bar q$), as in
the SM case, are gauge singlet under $SU(3)_L$ and carry a representation content given by 
 
\begin{align}
( d_R,  s_R, b_R)&\in (\bar 3, 1, 1/3)\\
( u_R,  c_R,  t_R) &\in(\bar 3, 1, -2/3)\\
( D_R,  S_R) &\in (\bar 3, 1, 4/3)\\
 T_R &\in (\bar 3, 1, -5/3).
\end{align}
These states are not sufficient to cancel the $SU(3)_L^3$ anomaly,
which requires extra states from the leptonic sector assigned
to the $\bar{3}$ representation of the same gauge group. 
The solution is a democratic arrangement of the
three lepton generations into triplets of $SU(3)_L$:
\begin{equation}
l=\left(
\begin{array}{c}
l_L\\
\nu_l\\
\bar l_R
\end{array}
\right),\quad l\in(1, \bar 3, 0),\quad l=e,\ \mu,\ \tau.
\end{equation}
In this way the  contribution of $Q_1$ (+9) and $Q_2$ (+9)
in the $3$ of $SU(3)_L$ to the $SU(3)_L^3$ anomaly is balanced by 
the one coming from $Q_3$ (-9) and by the three generations
of leptons $L_i$ $((-3)\times 3)$ assigned to the $\bar{3}$ 
representation of $SU(3)_L$.
For the $SU(3)_c^3$ anomaly, the cancellation is similar to the SM, with a complete balance between left-handed $(3 \times 3)$ colour triplets
and right-handed $(-(3 +3 +2 +1))$ color anti-triplets, and is 
henceforth obtained only within the quark sector.

This arrangement of the fermions corresponds to the version
of the model proposed in Refs.~\cite{PHF,PP}.
There are, however, other versions of similar models proposed more
recently that differ from the original formulation in some essential aspects. For instance, such variants affect the fermion sector with the addition of extra leptons and extra quarks \cite{buras331} or the flipping
between quarks and leptons \cite{gauld,flipp331} where all the quark fields are in the same representation of $SU(3)_L$ and the lepton families are in different ones, inverting the structure presented above. The non-universality of the underlying gauge structure is one of the most significant aspects of the model which deserves close attention, especially for its implications
on the flavour sector.

The scalars of the model, responsible for the electroweak symmetry breaking, come in three triplets of $SU(3)_L$.
\begin{equation}
\rho=\left(
\begin{array}{c}
\rho^{++}\\
\rho^+\\
\rho^0
\end{array}
\right)\in(1,3,1),\quad\eta=\left(
\begin{array}{c}
\eta^+\\
\eta^0\\
\eta^-
\end{array}
\right)\in(1,3,0),\quad\chi=\left(
\begin{array}{c}
\chi^0\\
\chi^-\\
\chi^{--}
\end{array}
\right)\in(1,3,-1).
\end{equation}
The breaking $SU(3)_L\times U(1)_X\to U(1)_{\rm{em}}$ is obtained in two steps. The vacuum expectation value (vev) of the neutral component of $\rho$ gives mass to the extra gauge bosons, $Z', Y^{++}$ and $Y^+$, as well as the extra quarks
$D$, $S$ and $T$. This causes the breaking from $SU(3)_L\times U(1)_X$ to $SU(2)_L\times U(1)_Y$; the usual spontaneous symmetry breaking
mechanism from $SU(2)_L\times U(1)_Y$ to $U(1)_{\rm{em}}$ is then
obtained through the vevs of the neutral components. The potential is then given by the expression 
\begin{align}
V&= m_1\, \rho^*\rho+m_2\,\eta^*\eta+m_3\,\chi^*\chi\nn\\
&\quad+\lambda_1 (\rho^*\rho)^2+\lambda_2(\eta^*\eta)^2+\lambda_3(\chi^*\chi)^2\nn\\
&\quad +\lambda_{12}\rho^*\rho\,\eta^*\eta+\lambda_{13}\rho^*\rho\,\chi^*\chi+\lambda_{23}\eta^*\eta\,\chi^*\chi\\
&\quad +\zeta_{12}\rho^*\eta\,\eta^*\rho+\zeta_{13}\rho^*\chi\,\chi^*\rho+\zeta_{23}\eta^*\chi\,\chi^*\eta\nn\\
&\quad +\sqrt2 f_{\rho\eta\chi} \rho\, \eta\, \chi\nn
\end{align}
and the neutral component of each triplet acquires a vev and is expanded as
\bea
\rho^0=\frac{1}{\sqrt2}v_\rho+\frac{1}{\sqrt2}\left(\rm{Re}\,\rho^0+i\,\rm{Im}\,\rho^0\right)\\
\eta^0=\frac{1}{\sqrt2}v_\eta+\frac{1}{\sqrt2}\left(\rm{Re}\,\eta^0+i\,\rm{Im}\,\eta^0\right)\\
\chi^0=\frac{1}{\sqrt2}v_\chi+\frac{1}{\sqrt2}\left(\rm{Re}\,\chi^0+i\,\rm{Im}\,\chi^0\right).
\eea
The minimization conditions of the potential, defined by $\frac{\partial V}{\partial\Phi}|_{\Phi=0}=0$, then take the form 
\bea
m_1v_\rho+\lambda_1v_\rho^3+\frac{\lambda_{12}}{2}v_\rho v_\eta^2-f_{\rho\eta\chi}v_\eta v_\chi+\frac{\lambda_{13}}{2}v_\rho v_\chi^2=0\\
m_2v_\eta+\lambda_2v_\eta^3+\frac{\lambda_{12}}{2}v_\rho^2 v_\eta-f_{\rho\eta\chi}v_\rho v_\chi+\frac{\lambda_{23}}{2}v_\eta v_\chi^2=0\\
m_3v_\chi+\lambda_3 v_\chi^3+\frac{\lambda_{13}}{2}v_\rho^2 v_\chi-f_{\rho\eta\chi}v_\rho v_\eta+\frac{\lambda_{23}}{2}v_\eta^2 v_\chi=0
\eea
and are solved for $m_1$, $m_2$ and $m_3$. After spontaneous symmetry breaking the gauge and the mass eigenstates of $\rho$, $\eta$ and $\chi$ are related by rotation matrices $\mathcal{R}$, whose expressions are rather lengthy to be given here. In the CP-even neutral sector, for instance, the transformation is given by
\bea
h_i=\mathcal{R}^S_{ij} H_j,
\eea
where $\vec H=(\rm{Re}\,\rho^0,\,\rm{Re}\,\eta^0,\,\rm{Re}\,\chi^0)$ and $\vec h=(h_1,\,h_2,\,h_3)$.
The explicit expression of the mass matrices of the scalar sector, both neutral and charged, can be computed quite directly
\bea
\mathcal{M}^h=\left(
\footnotesize\renewcommand*{\arraystretch}{1.5}
\begin{array}{ccc}
 2 \lambda_1 v_{\rho}^2+\frac{f_{\rho\eta\chi}}{v_{\rho}} V^2 \cos\beta \sin\beta\;, & \lambda_{12} v_{\rho} V \sin\beta-f_{\rho\eta\chi}
   V \cos\beta\;, & V (\lambda_{13} v_{\rho} \cos\beta-f_{\rho\eta\chi}
   \sin\beta) \\
 \lambda_{12} v_{\rho} V \sin\beta-f_{\rho\eta\chi} V \cos\beta\;, & 2
   \lambda_2 V^2 \sin ^2\beta+f_{\rho\eta\chi} v_{\rho} \cot\beta\;, & \lambda_{23}
   V^2 \cos\beta \sin\beta-f_{\rho\eta\chi} v_{\rho} \\
 V (\lambda_{13} v_{\rho} \cos\beta-f_{\rho\eta\chi} \sin\beta)\;, & \lambda_{23}
   V^2 \cos\beta \sin\beta-f_{\rho\eta\chi} v_{\rho} \;,& 2 \lambda_3 V^2
   \cos ^2\beta+f_{\rho\eta\chi} v_{\rho} \tan\beta \\
\end{array}
\right),\nn\\
\eea

\bea
\mathcal{M}^a=\left(
\footnotesize\renewcommand*{\arraystretch}{1.5}
\begin{array}{ccc}
 \frac{f_{\rho\eta\chi}}{v_\rho} V^2 \cos\beta \sin\beta & f_{\rho\eta\chi}
   V \cos\beta & f_{\rho\eta\chi} V \sin\beta \\
 f_{\rho\eta\chi} V \cos\beta & f_{\rho\eta\chi} v_\rho \cot\beta & f_{\rho\eta\chi}
   v_\rho \\
 f_{\rho\eta\chi} V \sin\beta & f_{\rho\eta\chi} v_\rho & f_{\rho\eta\chi} v_\rho \tan\beta \\
\end{array}
\right),
\eea

\bea
\hspace{-1.5cm}
\mathcal{M}^{h^\pm}=\left(
\scriptsize\renewcommand*{\arraystretch}{1.5}
\begin{array}{cccc}
 \frac{V^2}{2 v_\rho} \sin\beta (2 f_{\rho\eta\chi} \cos\beta+\zeta_{12} v_\rho \sin\beta) & f_{\rho\eta\chi} V \cos\beta+\frac{1}{2} \zeta_{12}
   v_\rho V \sin\beta & 0 & 0 \\
 f_{\rho\eta\chi} V \cos\beta+\frac{1}{2} \zeta_{12} v_\rho V \sin\beta & \frac{1}{2} v_\rho (\zeta_{12} v_\rho+2 f_{\rho\eta\chi} \cot\beta) & 0
   & 0 \\
 0 & 0 & \frac{1}{2} \zeta_{23} V^2 \cos ^2\beta+f_{\rho\eta\chi} v_\rho \cot\beta & \frac{1}{2} \zeta_{23} \cos\beta \sin\beta
   V^2+f_{\rho\eta\chi} v_\rho \\
 0 & 0 & \frac{1}{2} \zeta_{23} \cos\beta \sin\beta V^2+f_{\rho\eta\chi}
   v_\rho & \frac{1}{2} \zeta_{23} V^2 \sin ^2\beta+f_{\rho\eta\chi} v_\rho \tan\beta \\
\end{array}
\right),\nn\\
\eea

\bea
\mathcal{M}^{h^{\pm\pm}}=\left(
\footnotesize\renewcommand*{\arraystretch}{1.5}
\begin{array}{cc}
 \frac{V^2}{2 v_\rho} \cos\beta (\zeta_{13} v_\rho \cos\beta+2 f_{\rho\eta\chi} \sin\beta) & \frac{1}{2} \zeta_{13} v_\rho V \cos\beta+f_{\rho\eta\chi} V \sin\beta \\
 \frac{1}{2} \zeta_{13} v_\rho V \cos\beta+f_{\rho\eta\chi} V \sin\beta & \frac{1}{2} v_\rho (\zeta_{13} v_\rho+2 f_{\rho\eta\chi} \tan\beta) \\
\end{array}
\right),
\eea
while their diagonalization has been performed numerically with the parameter
choice which will be discussed below.
Notice that we have used the definition
\bea
V=\sqrt{v_\eta^2 + v_\chi^2}= 246 \;\rm{GeV}\ ,\  \tan\beta=\frac{v_\eta}{v_\chi}
\eea
for the vevs appearing in the equations above.

\section{Phenomenological Analysis}

In this section we wish to present a phenomenological analysis,
displaying possible signals of the bilepton model at the LHC.
As discussed in the previous sections, the striking feature of the
model is the prediction of doubly-charged gauge bosons
$Y^{++}$ and $Y^{--}$ and we shall investigate several physical
observables which will be sensitive to the existence of  bileptons.

\subsection{State of the art of bilepton phenomenology}

Before presenting our results, we would like to review
several interesting studies which
have been so far undertaken and overlap with our analysis
\cite{Dutta, Dion, Barreto1,Barreto2,Alves,nepo}.

Ref.~\cite{Dutta} studies single doubly-charged bilepton
$(Y^{++,--})$ production in association with
exotic quarks $D$ of charge $-4/3$, say $pp \to Y^{++}D$,
and possible subsequent decays, such as $D\to Y^{--} q$ or $D\to Y^-q'$,
$q$ and $q'$ being Standard Model quarks, followed by the
leptonic decays of $Y^{--}$ and $Y^-$.
The total inclusive cross section was then computed at leading-order (LO),
for both bilepton signal and $ZZ$ and $W^\pm Z$ backgrounds, at Tevatron
and LHC, as a function of $Y$ and exotic-quark masses.

The work in \cite{Dion} considered instead $Y^{++}Y^{--}$ pairs
in Drell--Yan 
processes, i.e. $pp\to Z,\gamma,Z'\to Y^{++}Y^{--}$, as well as the
production of pairs of one doubly- and one singly-charged bilepton, i.e.
$Y^{++}Y^-(Y^{--}Y^+)$,
mediated by $Z$, $Z'$, $W^\pm$ and photons.
The total cross section was computed at LO for different values of 
$Y$ and $Z'$ masses, for both vector and scalar
bileptons. The potential for bilepton discovery at Tevatron and LHC was
discussed as well.

Ref.~\cite{Barreto1} investigated
the pair production of singly-charged heavy vectors, which the authors
call $V$, say $pp \to V^+V^-$, the dependence of the LO cross section
on the mediating-$Z'$ mass and its comparison with respect to the
$W^+W^-$-driven Standard Model background. 

The analysis in \cite{Barreto1} was then extended in \cite{Barreto2}
with the inclusion of the leptonic
decays of singly-charged bileptons and the comparison of
few leptonic
distributions against the
$WW$ background. In \cite{Barreto2} the doubly-charged Drell--Yan
bilepton production was also accounted for and
same-sign dilepton invariant mass and
transverse momentum spectra
were presented, in terms of $Z'$ and $Y$ masses,
at LO matrix-element level;
no background was nevertheless computed.

Ref.~\cite{Alves} investigated the production of 
leptoquarks, labelled as $J_3$, in $pp$ collisions, through
both electroweak (Drell--Yan like) and strong interactions, and possible
decays of $J_3$ into singly- or doubly-charged bileptons, 
denoted by $V^\pm$ and $U^{++,--}$,
plus a bottom or a top quark.
The total cross sections were calculated at LO and
the full process $pp\to J_3 J_3 \to (Y^{++}Y^{--}) (b \bar{b})\to 
(\ell^+\ell^+) (\ell^-\ell^-) (b\bar b)$
was then examined; the same-sign
$\ell\ell$ and $\ell\ell b$
invariant masses were then studied, varying $m_{J_3}$ and $m_{Z'}$.
Once again, the results in \cite{Alves} were just at matrix-element
level, with no parton shower or hadronization matching;
the background distributions were not shown, as the authors
said that they were negligible with respect to the signal,
after applying acceptance cuts on jets (partons) and leptons.

More recently, Ref.~\cite{nepo} explored the LHC potential
for discovering bileptons in $pp\to Y^{++}Y^{--}\to\mu^+\mu^+\mu^-\mu^-$,
with $Y$-pair production mediated by either a
vector boson ($\gamma$, $Z$, $Z'$) or an leptoquark $Q$.
The authors of \cite{nepo} implemented the bilepton model
in the matrix-element generator \texttt{CalcHEP}
\cite{calchep}, matched with \texttt{PYTHIA} \cite{pythia}, and 
found that
the 7 TeV ATLAS data
on doubly-charged Higgs production
\cite{aad}, for a luminosity $\mathcal{L}=5~{\rm fb}^{-1}$,
allow one to exclude bileptons between 250 and
500 GeV, depending on the mass of the exotic quarks.
Such results were extended to $\sqrt{s}=13$~TeV and
$\mathcal{L}=50~{\rm fb}^{-1}$: by means of a single-bin analysis,
based on a Bayesan technique, 
lower bound $m_{Y^{\pm,\pm}}>850$~GeV was obtained.

The analysis which we shall carry out will be concentrated on events
with two same-sign lepton pairs and two jets and it should
complement the existing
studies on bilepton phenomenology.
In fact, as discussed above, 
the work in Refs.~\cite{Dutta, Dion, Barreto1,Barreto2,Alves,nepo}
did cover a large portion of the relevant parameter space for bilepton
production at LHC.
In particular, Ref.~\cite{Alves}, where
final states with two jets and two bileptons were explored,
determined the LHC reach for bilepton discovery 
at both 7 and 14 TeV energies, for a wide range of $Z'$ and
leptoquark masses. 

On the contrary, in the present paper 
we shall undertake our analysis at 13 TeV and limit ourselves to
a benchmark point of the
parameter space (mass spectrum),
chosen in such a way to enhance the bilepton signal
and consistent with the Higgs discovery as well as the present exclusion limits
on new particles, such as $Y$ and $Z'$.
Moreover, as done in \cite{nepo} for jetless $YY$-pair production,
we shall implement our model in the framework of a full
Monte Carlo simulation,
including parton showers and hadronization.
The UFO (Universal FeynRules Output) model generated by \texttt{SARAH}
will be used by  \texttt{MadGraph} to generate the relevant amplitudes,
which will be matched to \texttt{HERWIG} for parton showers and
hadronization.
Thanks to the implementation of the model in \texttt{MadGraph},
the matrix elements for any process, such as
$pp\to  Y^{++}Y^{--}jj$, will include all the possible subprocesses 
predicted by the model under investigation.

Although the total cross section will still be the LO one,
the differential distributions will account
for multi-parton radiation and will therefore be equivalent to those
yielded by a leading-logarithmic resummed calculation
(see, e.g., \cite{cmw} on the comparison between parton shower algorithms
and resummations). The matching of matrix elements and parton showers will
also enrich the jet substructure, thus allowing us to implement an actual
jet-clustering algorithm: on the contrary, in the work \cite{Alves}, since
there was no showering, jets were just identified as partons at the
amplitude (\texttt{MadGraph}) level.
Furthermore, our Monte Carlo event generation has been designed in such a way
that it can be directly interfaced to any detector simulation, so that it 
can be straightforwardly employed by the ATLAS and CMS Collaborations to search
for doubly-charged
bileptons, extending the current analyses on doubly-charged scalar
Higgs bosons \cite{aad,hh1,hh2}.

The same procedure adopted for the bilepton signal will be followed
for the Standard Model backgrounds, which will be simulated
by \texttt{MadGraph} and matched to \texttt{HERWIG} for showers and
hadronization. Therefore, unlike Refs.~\cite{Barreto2,Alves},
which did include the background, but only at amplitude-level,
even our Standard Model distributions will
account for multiple gluon/quark radiation from initial- and final-state
partons and jet-clustering algorithm implementation.

\begin{figure}[t]
\centering
\mbox{\subfigure[]{
\includegraphics[width=0.2\textwidth]{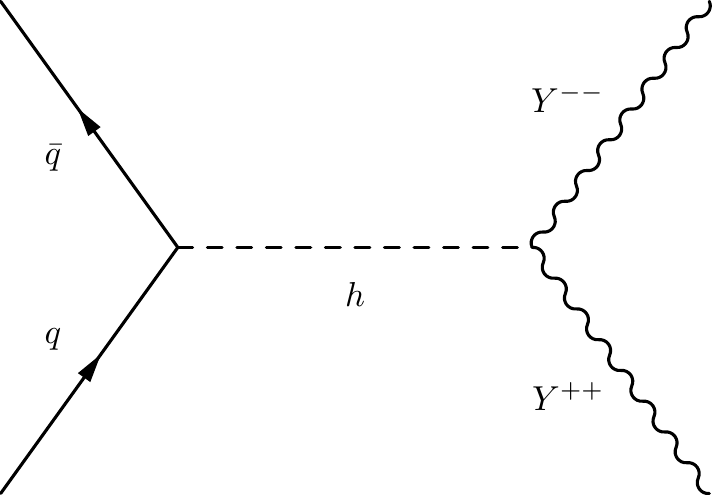}}\hspace{.55cm}
\subfigure[]{\includegraphics[width=0.2\textwidth]{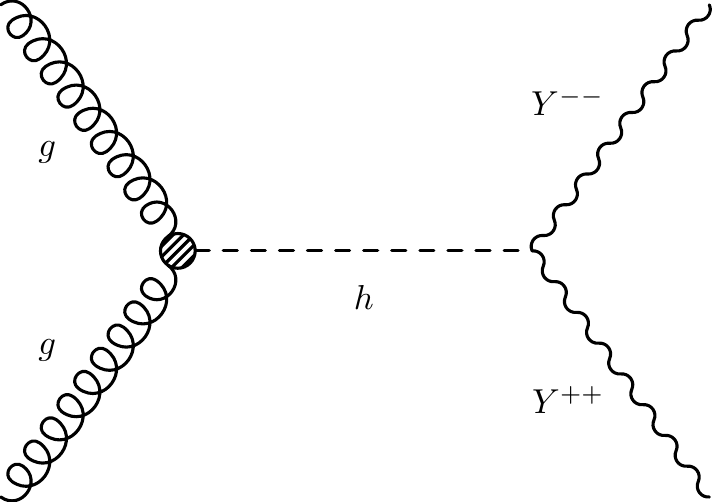}}}
\mbox{\subfigure[]{
\includegraphics[width=0.2\textwidth]{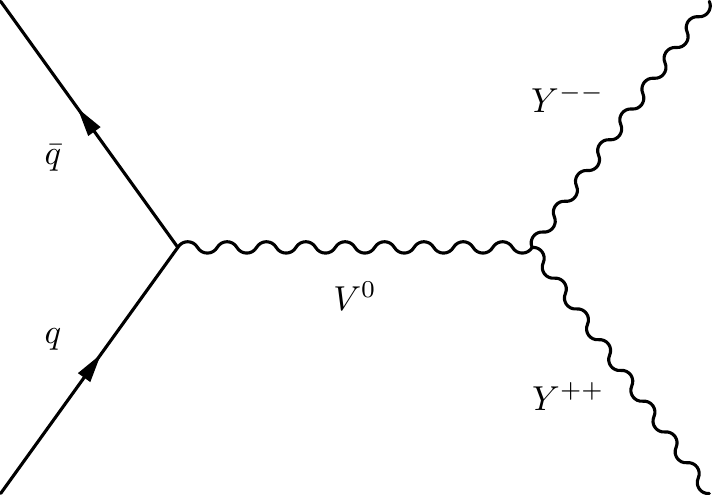}}\hspace{.55cm}
\subfigure[]{\includegraphics[width=0.2\textwidth]{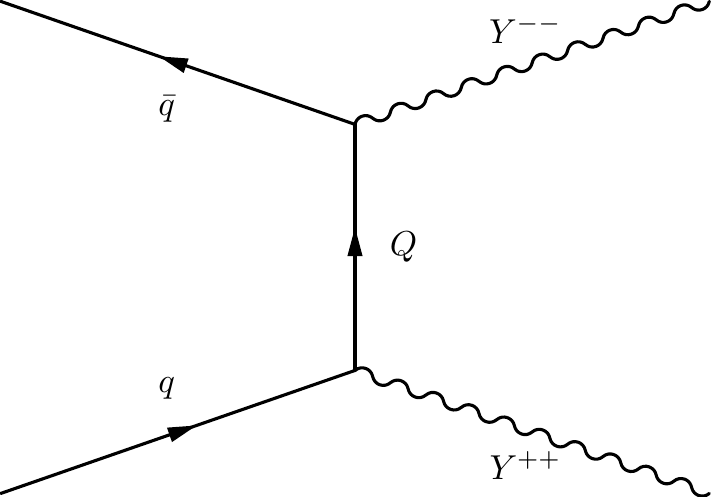}}}
\caption{List of the typical contributions to the events with two bileptons in the final state with no jets.}
 \label{jetless}
\end{figure}

Fig.~\ref{jetless} presents typical diagrams wherein bilepton
pairs $Y^{++}Y^{--}$ are produced at hadron colliders.
Figs.~\ref{jetless} (a) and \ref{jetless}(c) are Drell--Yan like
processes, where $h$ is the SM-like Higgs and $V^0$ a photon
or a $Z$ boson, whereas Fig.~\ref{jetless} (b) refers to bilepton-pair
production by gluon fusion, mediated by a $h$. In Fig.~\ref{jetless} (d)
$Y^{++}Y^{--}$ production occurs via the exchange of an exotic quark $Q$
in the $t$-channel.

\begin{figure}[t]
\centering
\mbox{\subfigure[]{
\includegraphics[width=0.2\textwidth]{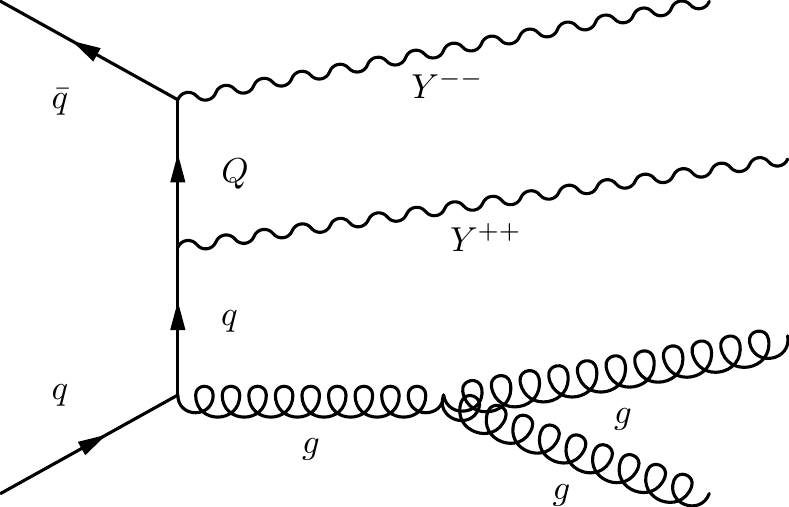}}
\subfigure[]{\includegraphics[width=0.2\textwidth]{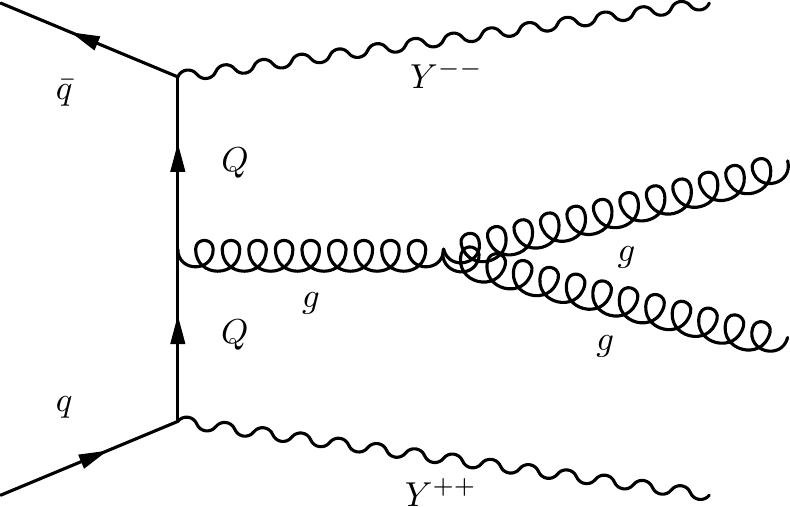}}
\subfigure[]{\includegraphics[width=0.2\textwidth]{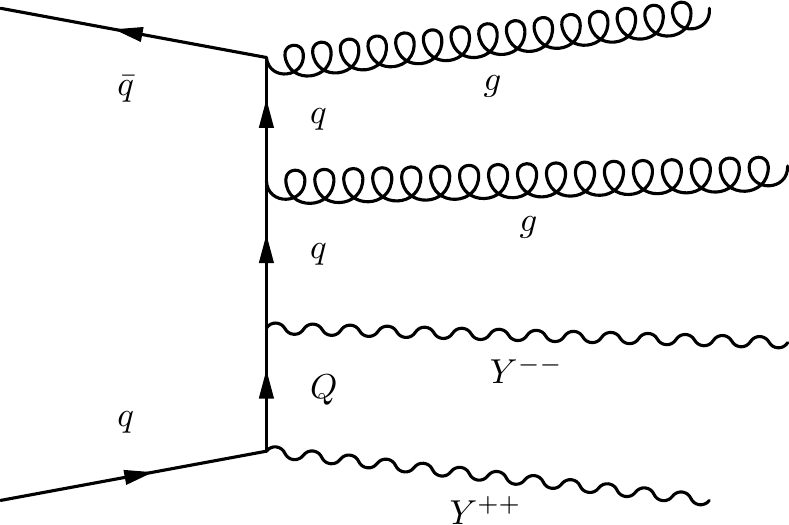}
}}
\mbox{\subfigure[]{
\includegraphics[width=0.2\textwidth]{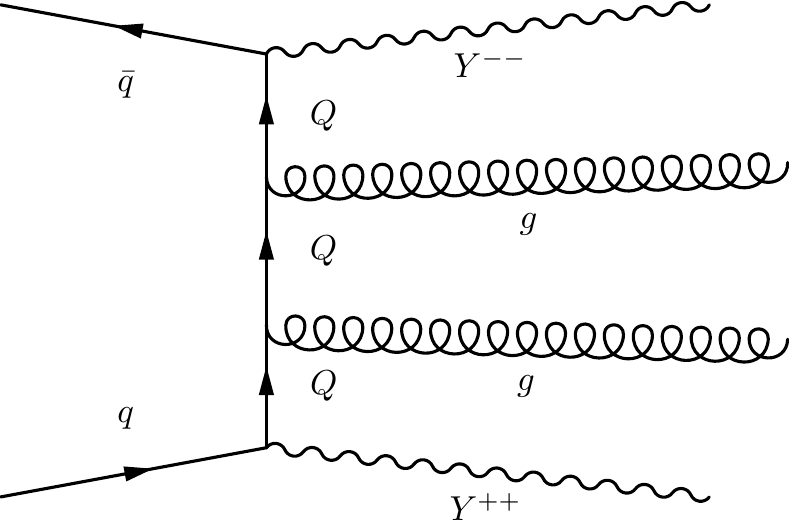}}
\subfigure[]{\includegraphics[width=0.2\textwidth]{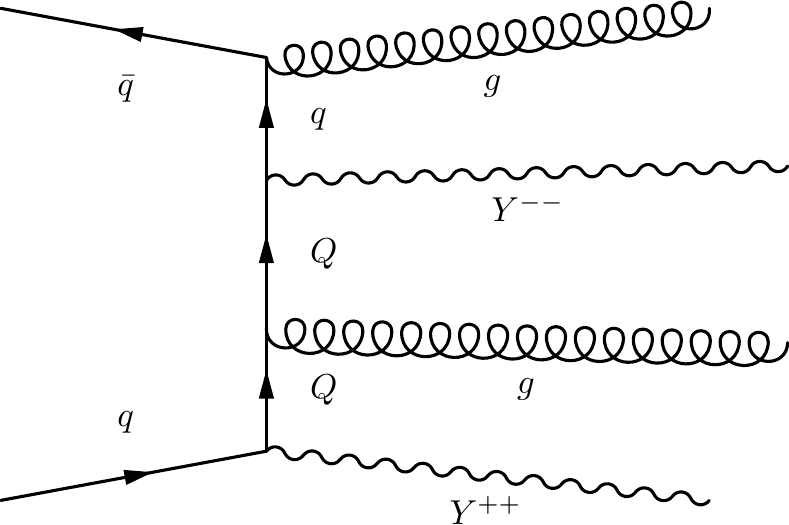}}
\subfigure[]{\includegraphics[width=0.2\textwidth]{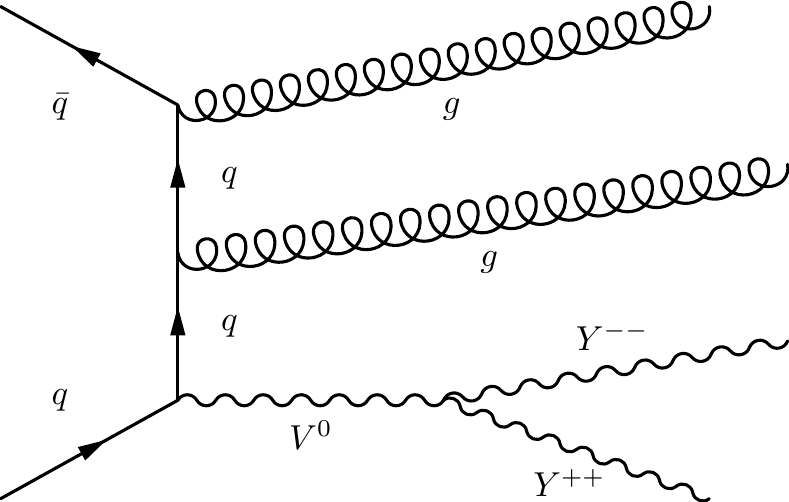}
}}
\caption{Typical contributions to the events with two bileptons in the final state at ${\cal O} (e^4 g^2)$ mediated by one or more exotic intermediate quarks with no extra $(W,\gamma, Z, Z')$ gauge bosons (diagrams $(a)-(e)$)
  and with one extra neutral gauge boson (diagram $(f)$).}
 \label{feyn1}
\end{figure}

\begin{figure}[t]
\centering
\mbox{\subfigure[]{
\includegraphics[width=0.2\textwidth]{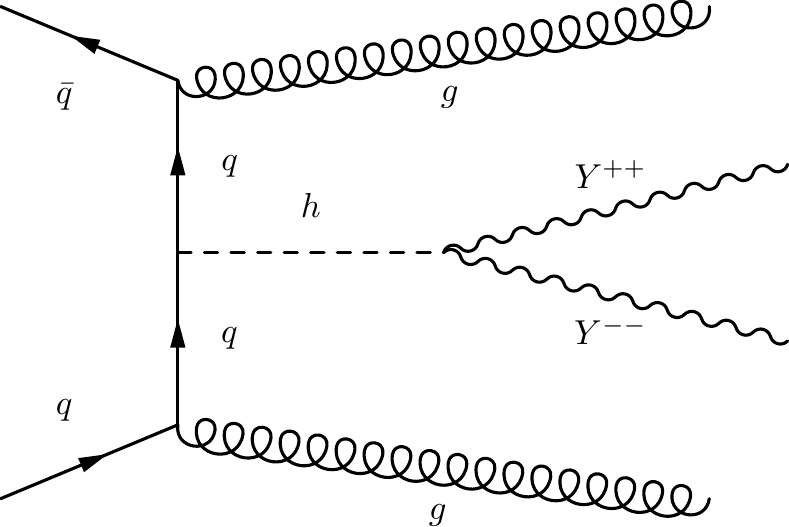}}\hspace{.75cm}
\subfigure[]{\includegraphics[width=0.2\textwidth]{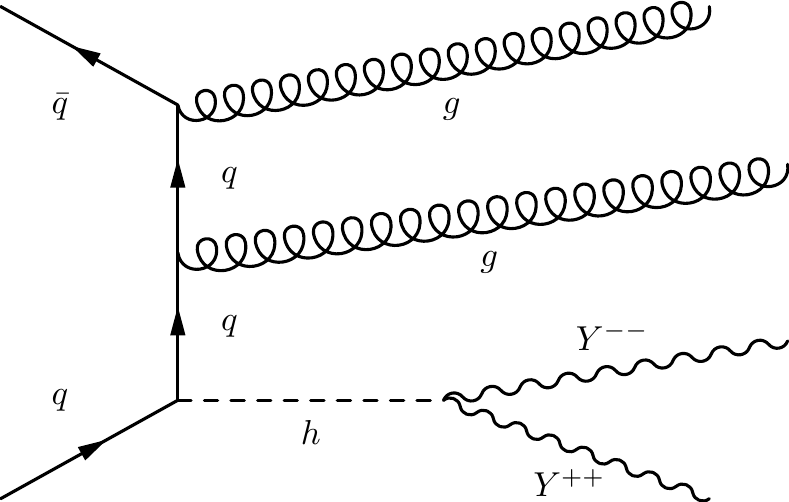}}}
\mbox{\subfigure[]{
\includegraphics[width=0.2\textwidth]{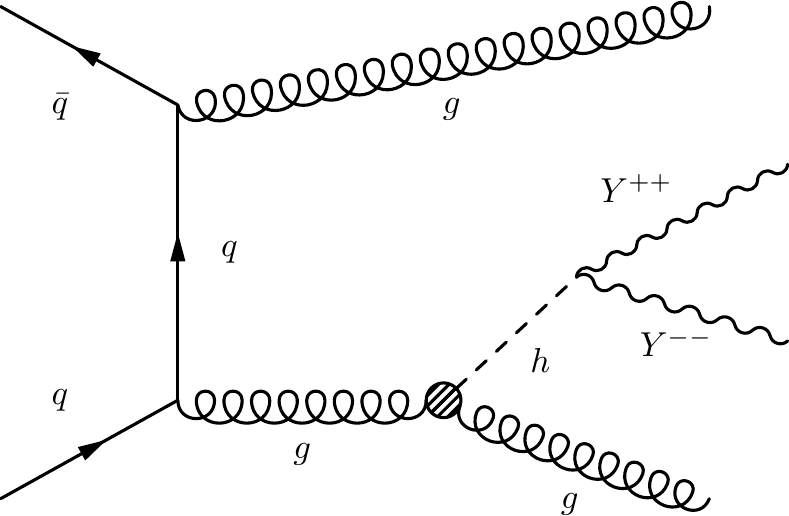}}\hspace{.75cm}
\subfigure[]{\includegraphics[width=0.2\textwidth]{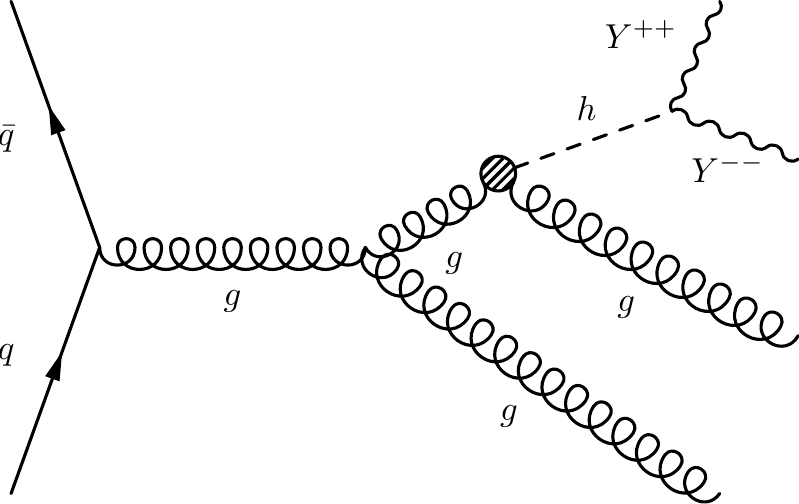}}}
\caption{Processes with two bileptons in the final state at ${\cal O}
  (e^4 g^2)$ with an intermediate Higgs scalar.}
 \label{feynh1}
\end{figure}

\begin{figure}[t]
\centering\mbox{\subfigure[]{
\includegraphics[width=0.2\textwidth]{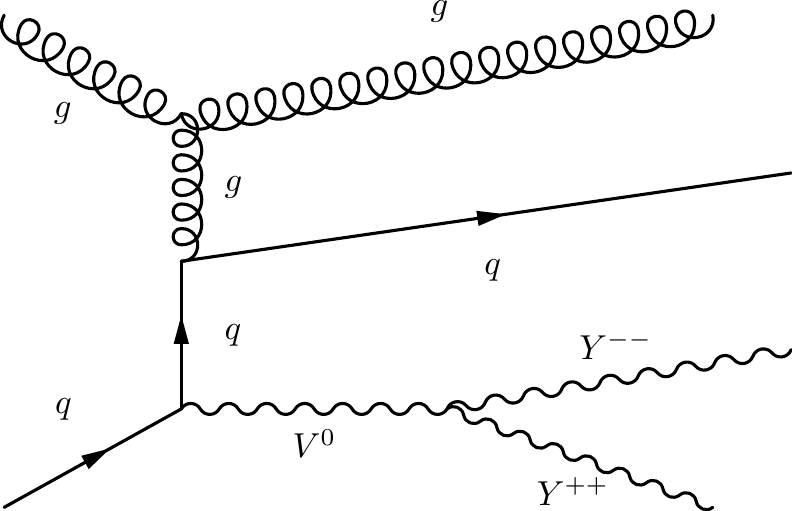}}
\subfigure[]{\includegraphics[width=0.2\textwidth]{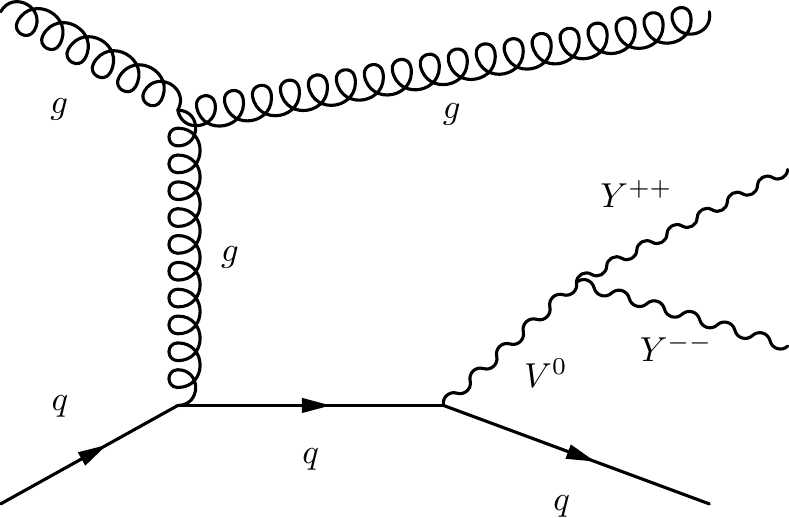}}
\subfigure[]{\includegraphics[width=0.2\textwidth]{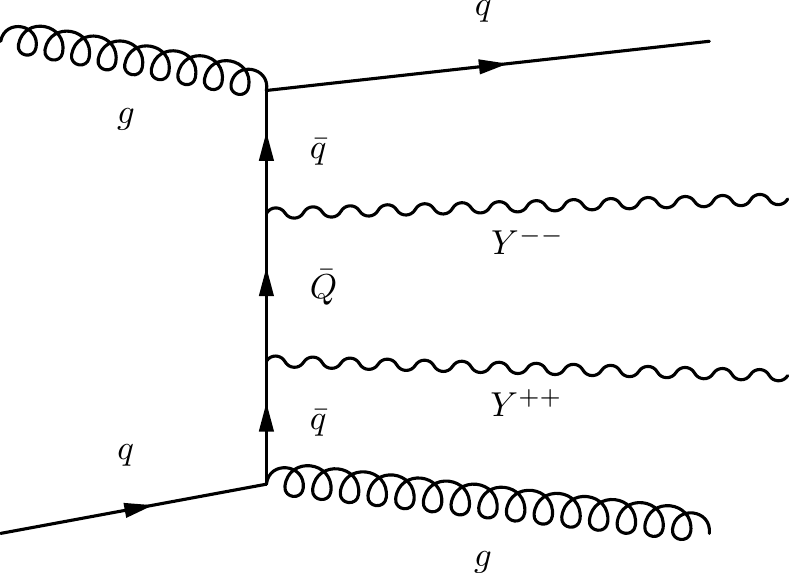}
}}
\mbox{\subfigure[]{
\includegraphics[width=0.2\textwidth]{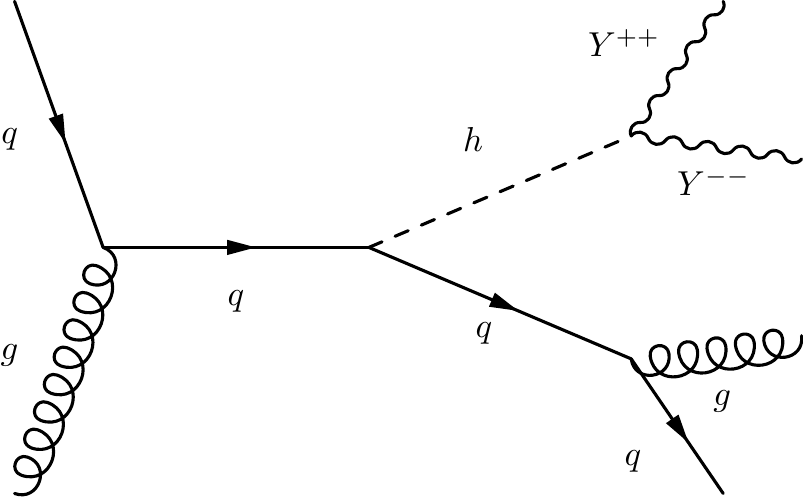}}
\subfigure[]{\includegraphics[width=0.2\textwidth]{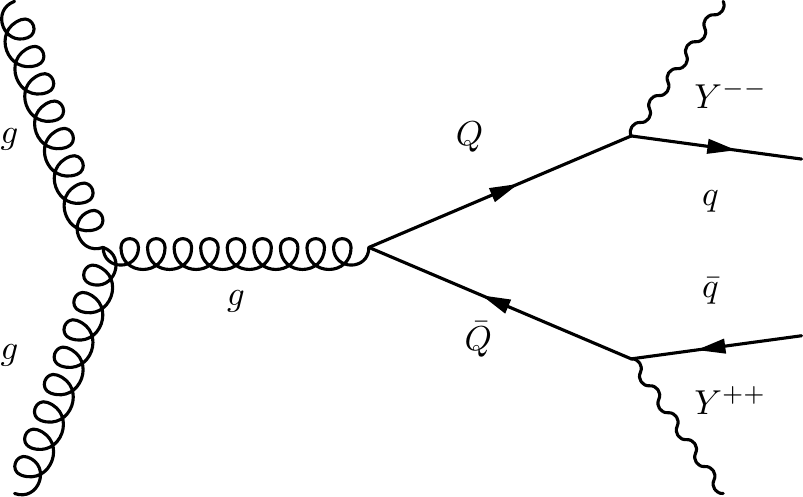}}
\subfigure[]{\includegraphics[width=0.2\textwidth]{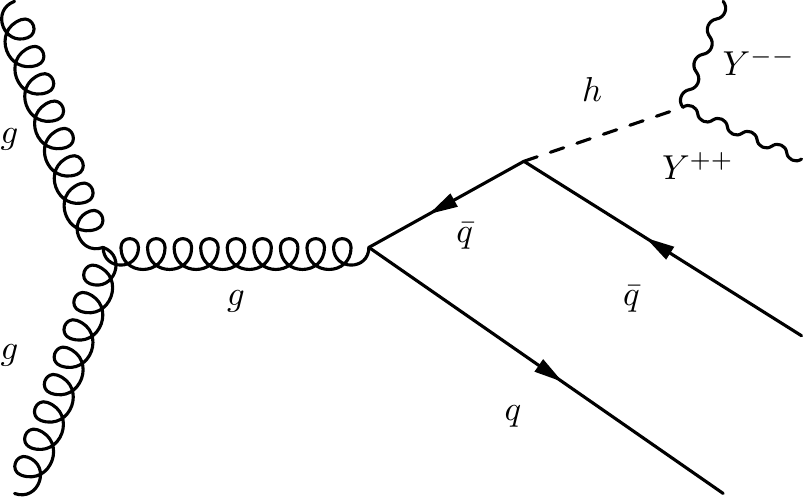}
}}
\mbox{\subfigure[]{
\includegraphics[width=0.2\textwidth]{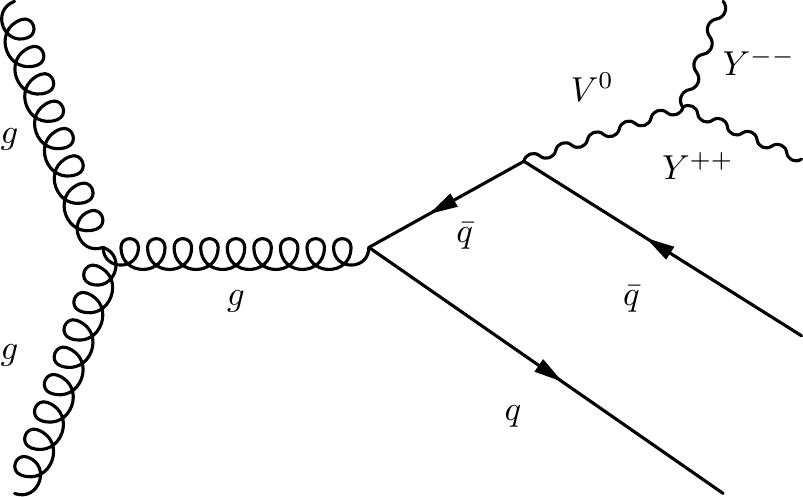}}}
\caption{Bilepton signal from quark-gluon fusion and gluon-gluon fusion.}
\label{qg1}
\end{figure}

In our study we shall nevertheless investigate
more exclusive final states, wherein the two same-sign lepton
pairs are accompanied by at least two additional jets ($jj$):
Some typical Feynman diagrams contributing to
final states with $Y^{++}Y^{--}jj$, where $j=q,g$ at parton level,
are depicted in Figs.~\ref{feyn1}-\ref{qg1}.
In detail, Fig.~\ref{feyn1} includes processes initiated by a $q\bar q$ pair
and mediated by light or exotic quarks, neutral vector bosons
or gluons, but no scalars. Fig.~\ref{feynh1} presents 
characteristic amplitudes where bilepton production occurs through
$h\to Y^{++}Y^{--}$ decays in processes with initial-state quarks. Finally, Fig.~\ref{qg1} shows instead some typical diagrams with
quark-gluon or gluon-gluon fusion in the initial state.
Although the $Z'\to Y^{++}Y^{--}$ subprocess still contributes
to  $Y^{++}Y^{--}jj$ final states, it is less crucial than in the Drell--Yan
processes investigated in \cite{Dion,Barreto1,Barreto2}. We can therefore
anticipate that we shall not present results for different
$Z'$-mass values, but we shall stick to one benchmark scenario.
Moreover, while the investigation in \cite{Alves} was based 
on $b$-flavoured jets, we shall account for jets of all types, initiated
by both light and heavy quarks, as well as by gluons.

\subsection{Results at 13 TeV LHC}
The phenomenology of our model is very rich due to the structure of 
both gauge and scalar sectors. For the sake of undertaking an actual
LHC analysis, possibly useful for the experimental BSM searches,
we have selected a specific benchmark point in parameter space, as given in Table \ref{BP}
and satisfying a certain number of constraints.
\begin{table}
\begin{center}
\renewcommand{\arraystretch}{1.4}
\begin{tabular}{|c|c|c|}
\hline\hline
\multicolumn{3}{|c|}{Benchmark Point}\\
\hline
\hline
$m_{h_1}=125.1$ GeV & $m_{h_2}=3172$ GeV& $m_{h_3}=3610$ GeV\\
\hline
$m_{a_1}=3595$ GeV&\multicolumn{2}{c}{}\\
\cline{1-2}
$m_{h^\pm_1}=1857$ GeV& $m_{h^\pm_2}=3590$ GeV& \multicolumn{1}{c}{}\\
\cline{1-2}
$m_{h^{\pm\pm}_1}=3734$ GeV&\multicolumn{2}{c}{}\\
\cline{1-2}
$m_{Y^{\pm\pm}}=873.3$ GeV& $m_{Y^\pm}=875.7$ GeV& \multicolumn{1}{c}{}\\
\cline{1-2}
$m_{Z'}=3229$ GeV&\multicolumn{2}{c}{}\\
\hline
$m_D=1650$ GeV & $m_S=1660$ GeV& $m_T=1700$ GeV\\
\hline
\hline
\end{tabular}
\caption{Benchmark point for a collider study consistent with the $\sim 125$ GeV Higgs mass.}\label{BP}
\end{center}
\end{table}\par
As already explained previously, the vevs of $\eta$ and $\chi$ are
responsible for the masses of the known quarks as well as for the masses of the $W$ and $Z$ gauge bosons.
Concerning the scalar sector, we require the lightest scalar
boson $h_1$ to be the candidate Higgs boson.
Our reference point has therefore been chosen in such a way to yield
$m_{h_1}\simeq 125\;\rm{GeV}$. Furthermore, 
the coupling of $h_1$ to $Z$ and $W$ has been chosen to be
SM-like:
\bea
\left|\frac{g_{h_1 ZZ}}{g^{\rm{SM}}_{hZZ}}\right| = 1.0 \pm 0.1     \\
\left|\frac{g_{h_1 WW}}{g^{\rm{SM}}_{hWW}}\right| = 1.0 \pm 0.1
\eea
The coupling of the Higgs boson $h_1$ to top quarks is also SM-like,
in such a way that our model 
reproduces the experimental cross section of Higgs production in gluon
fusion, i.e. $gg\to h$. 

Beside the SM-like Higgs, we have imposed some other constraints
on the new particles of the model.
As for bileptons, their mass at tree-level is given by
\bea
m_{Y^{\pm\pm}} = \frac{1}{2}g_2 \sqrt{v_\rho^2+V^2\cos^2\beta}.
\eea
In our benchmark point, we have set 
$m_{Y^{\pm\pm}}\simeq 873$~GeV, 
while the singly-charged $Y^{\pm}$ are
slightly heavier, i.e. $m_{Y^{\pm}}\simeq 876$~GeV;
such mass values are consistent with the exclusion limits
obtained in Ref.~\cite{nepo}.
The masses of the extra quarks are related, of course, to
their Yukawa couplings. However, we have chosen $m_D$, $m_S$ and $m_T$ in such a way that the two-body decay of the extra neutral boson $Z'$ becomes kinematically forbidden, i.e. $m_{Z'}<2\, m_Q$.
We give in the Appendix the list of the relevant vertices of our model
that have been implemented.

In the following, we shall present results for the production of two
bileptons plus two jets at the LHC
\be
pp\to Y^{++}Y^{--}jj\to (\ell^+\ell^+)(\ell^-\ell^-)jj,
\label{signal}
\ee
where $\ell=e,\mu$.
The amplitude of process (\ref{signal}) is generated by 
the \texttt{MadGraph} code, matched with \texttt{HERWIG} 6
for shower and hadronization.
We have
set $\sqrt{s}=13$~TeV and chosen the
NNPDFLO1 parton distributions \cite{nnpdf}, which are the default sets in
\texttt{MadGraph}.

We cluster jets at parton level 
according to the $k_T$ algorithm \cite{kt} with $R=1$,
setting the following cuts
on jets and lepton transverse momentum ($p_T$),
pseudorapidity ($\eta$) and invariant opening angle ($\Delta R$):
\be\label{cuts}
p_{T,j}>30~{\rm GeV}, p_{T,\ell}>20~{\rm GeV}, |\eta_j|<4.5,
|\eta_\ell|<2.5, \Delta R_{jj}>0.4,  \Delta R_{\ell\ell}>0.1,
\Delta R_{j\ell}>0.4.
\ee
After such cuts are applied, the LO cross section, computed by 
\texttt{MadGraph}, reads
\be\sigma(pp\to YYjj\to 4\ell jj)\simeq
3.7~{\rm fb}.\ee

As for the background, final states with
four charged leptons and two jets may occur through intermediate
$Z$-boson pairs
\be
pp\to ZZ\ jj\to (\ell^+\ell^-)(\ell^+\ell^-)jj.
\label{zz}
\ee
We shall also consider processes with intermediate $t\bar tZ$ states,
with the $Z$'s and the
$W$'s from top quarks decaying leptonically, and require
a cut $\rm{MET}<100$~GeV on the missing transverse
energy carried by the neutrinos
\be
pp\to t\bar tZ\to (j\ell^+\nu_\ell)(j\ell^-\bar\nu_\ell)(\ell^+\ell^-).
\label{ttbar}
\ee
Obviously, in process (\ref{ttbar}) the two jets are to be considered
$b$-jets.
On the leptons and the jets of both background processes 
we set the same cuts as in (\ref{cuts}). The LO cross sections are then given by
\be
\sigma(pp\to ZZ\ jj\to 4\ell jj)\simeq 6.4~{\rm fb}\ ,\ 
\sigma(pp\to t\bar t Z\ jj\to 4\ell\  2\nu\  jj)\simeq 8.6~{\rm fb}.
\ee
In principle, within the backgrounds, one should also consider $t\bar th$
processes, with $h\to\ell^+\ell^-$. However, because of the tiny
coupling of the Higgs boson to electrons and muons, such a background turns
out to be negligible.

\begin{figure}[p]
\centering
\mbox{\subfigure[]{
\includegraphics[width=0.4\textwidth]{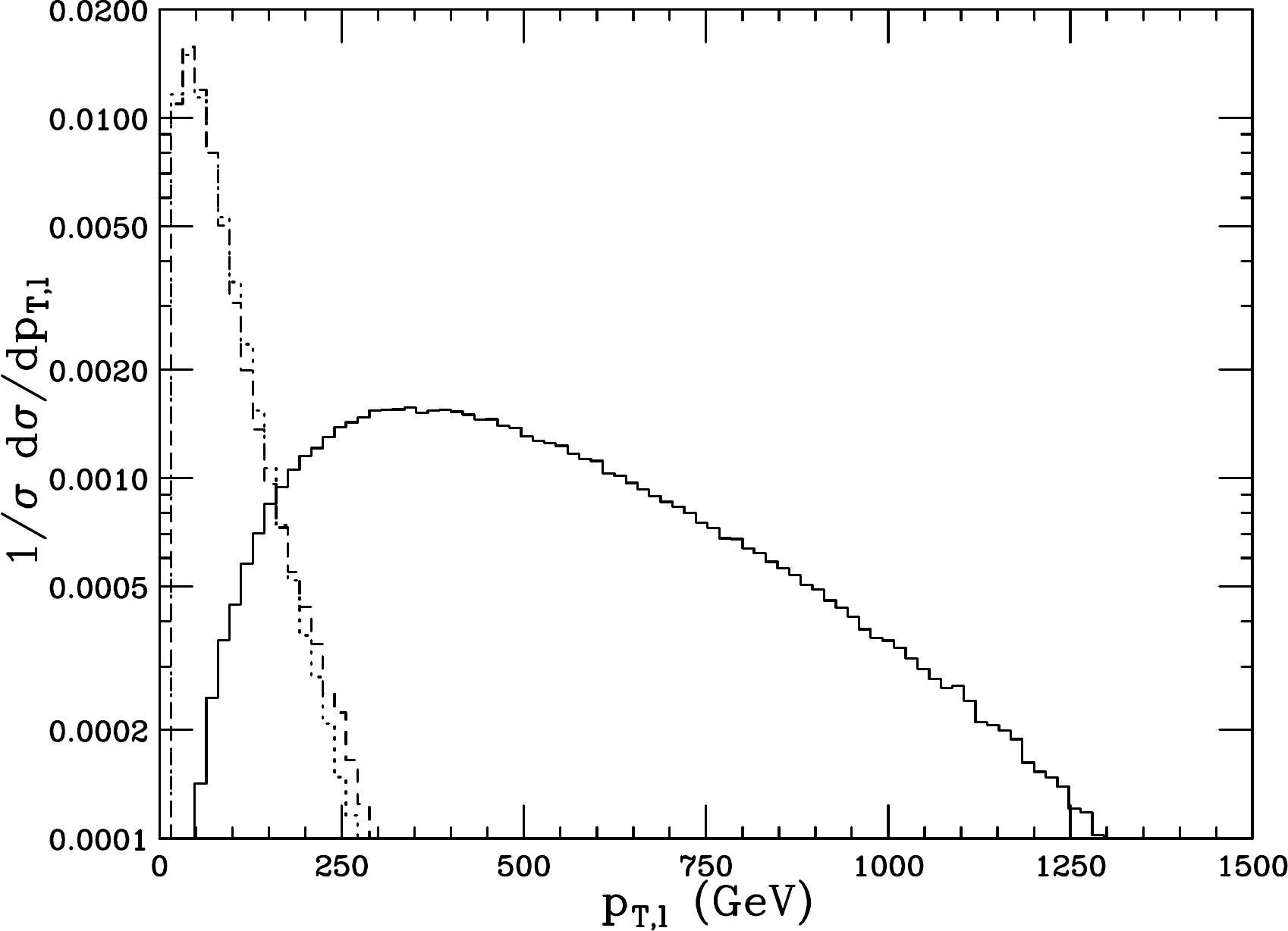}}\hspace{.55cm}
\subfigure[]{\includegraphics[width=0.4\textwidth]{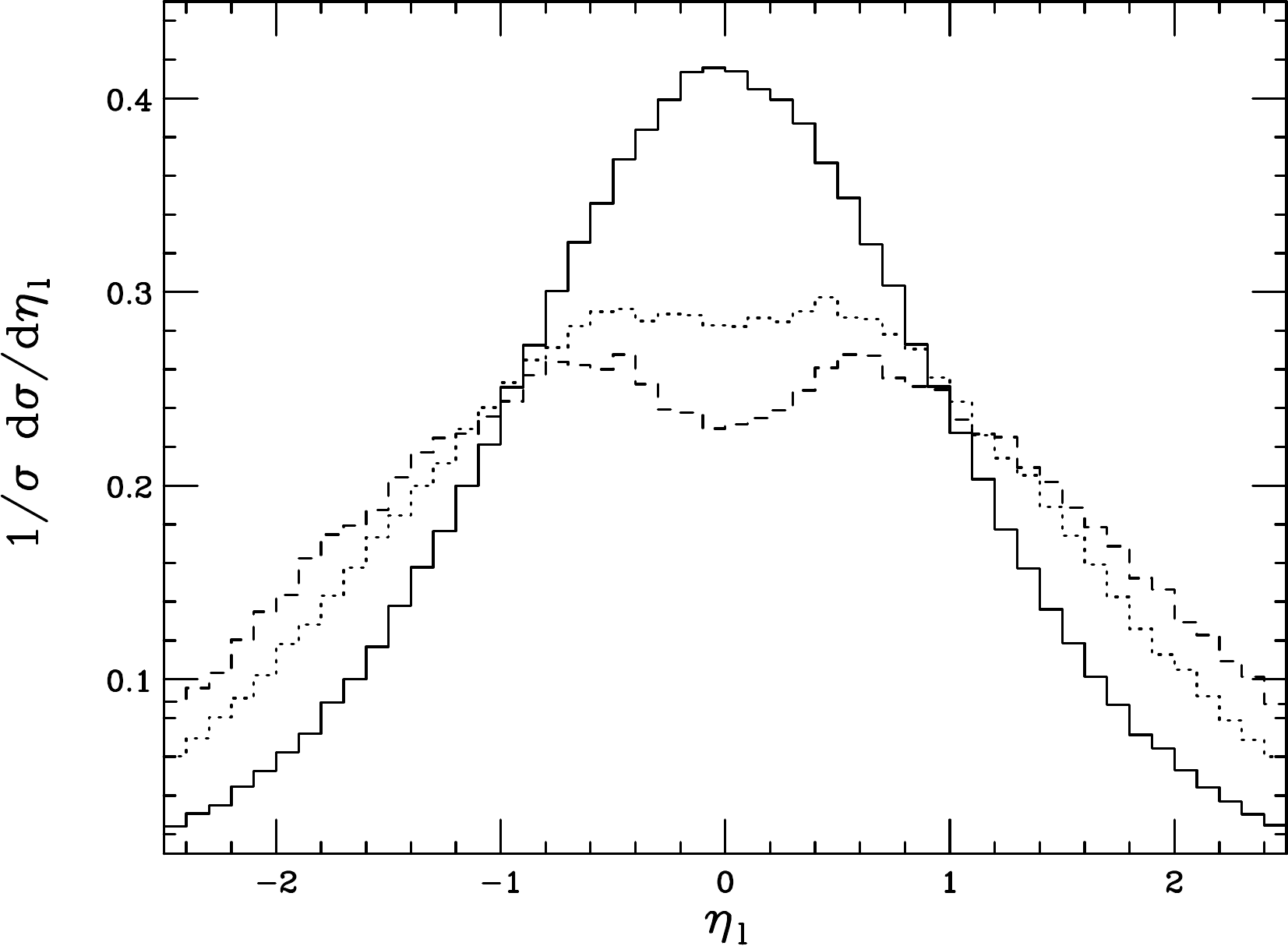}}}
\mbox{\subfigure[]{
\includegraphics[width=0.4\textwidth]{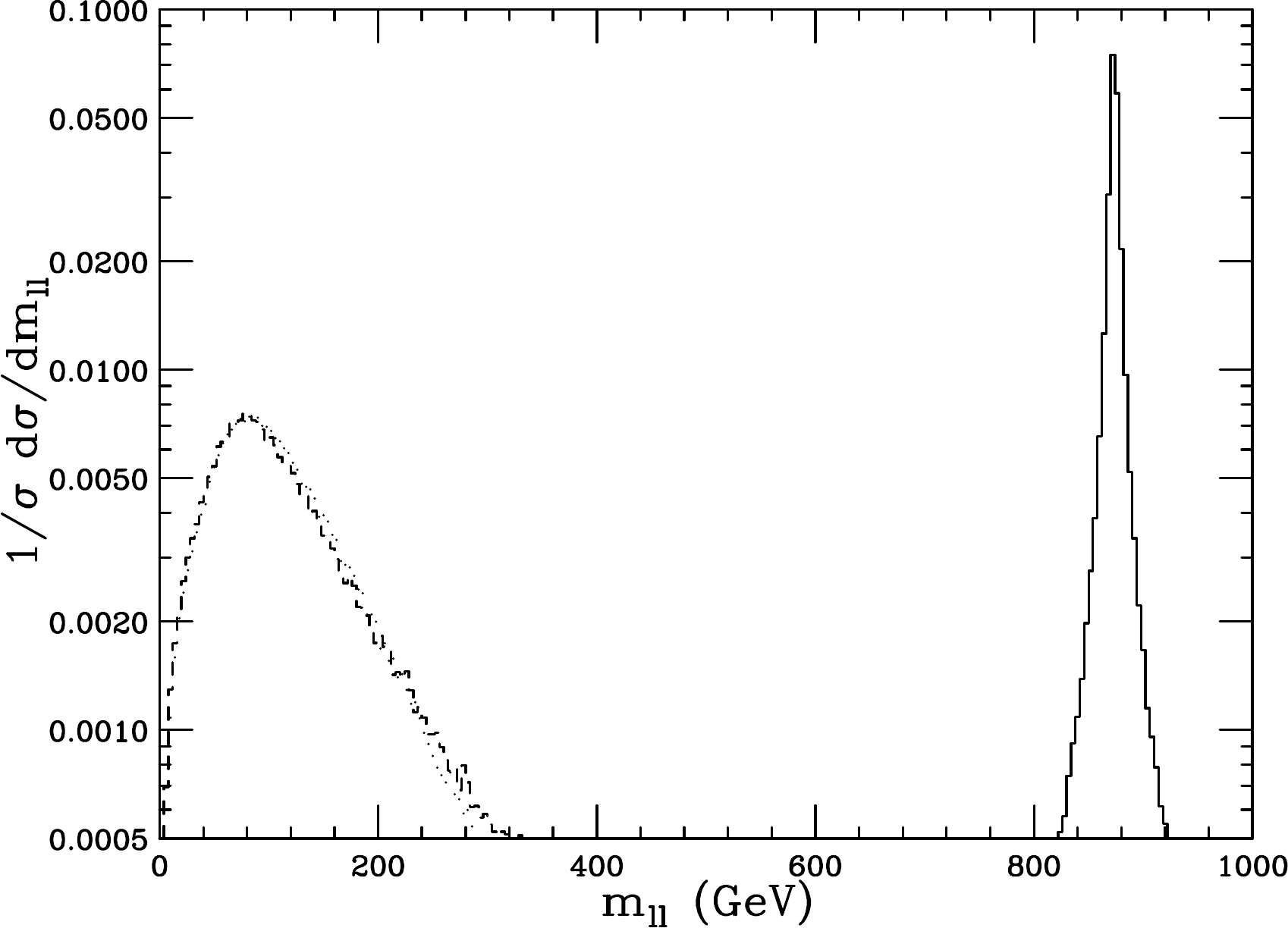}}\hspace{.55cm}
\subfigure[]{\includegraphics[width=0.4\textwidth]{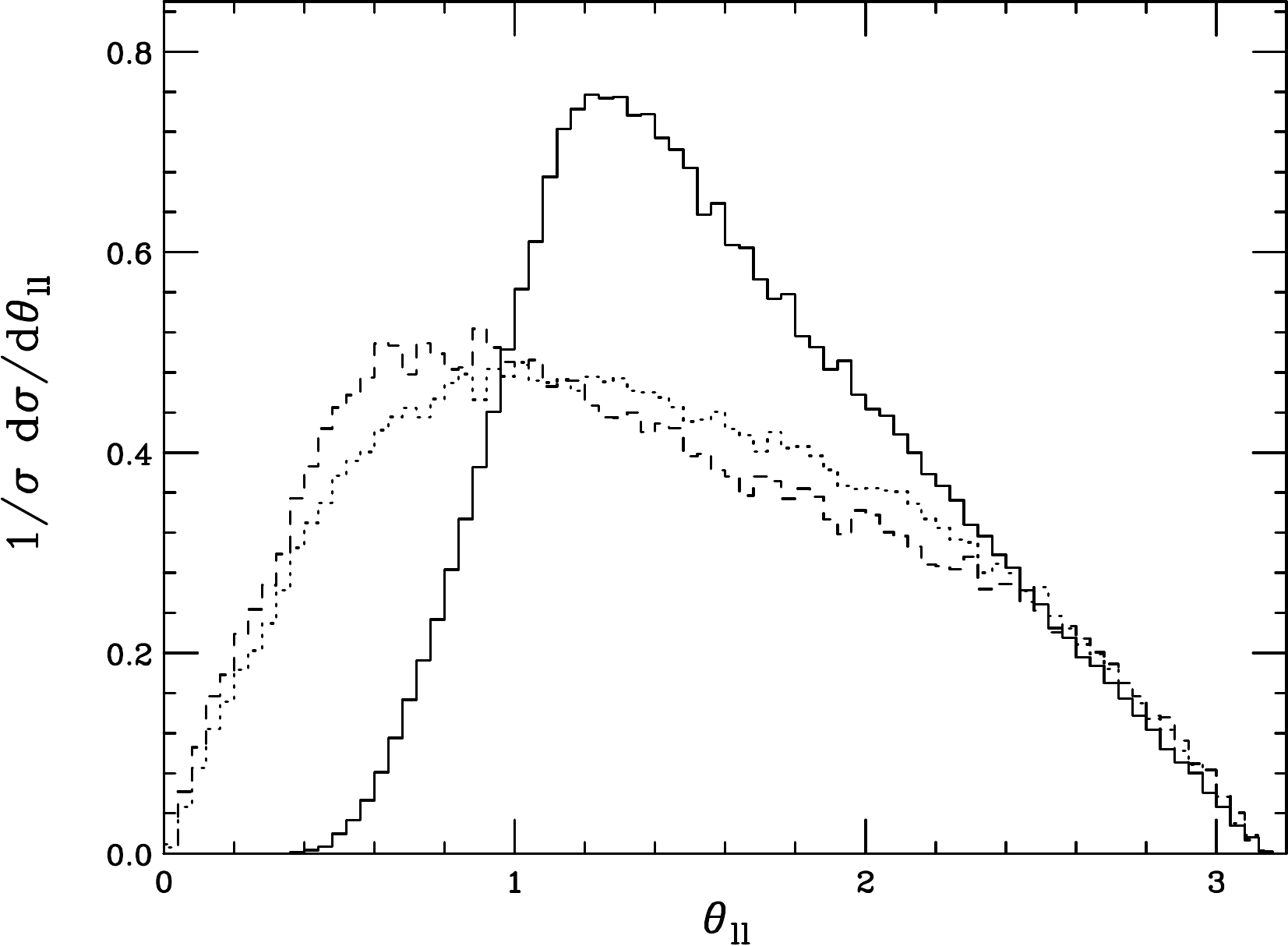}}}
\mbox{\subfigure[]{
\includegraphics[width=0.4\textwidth]{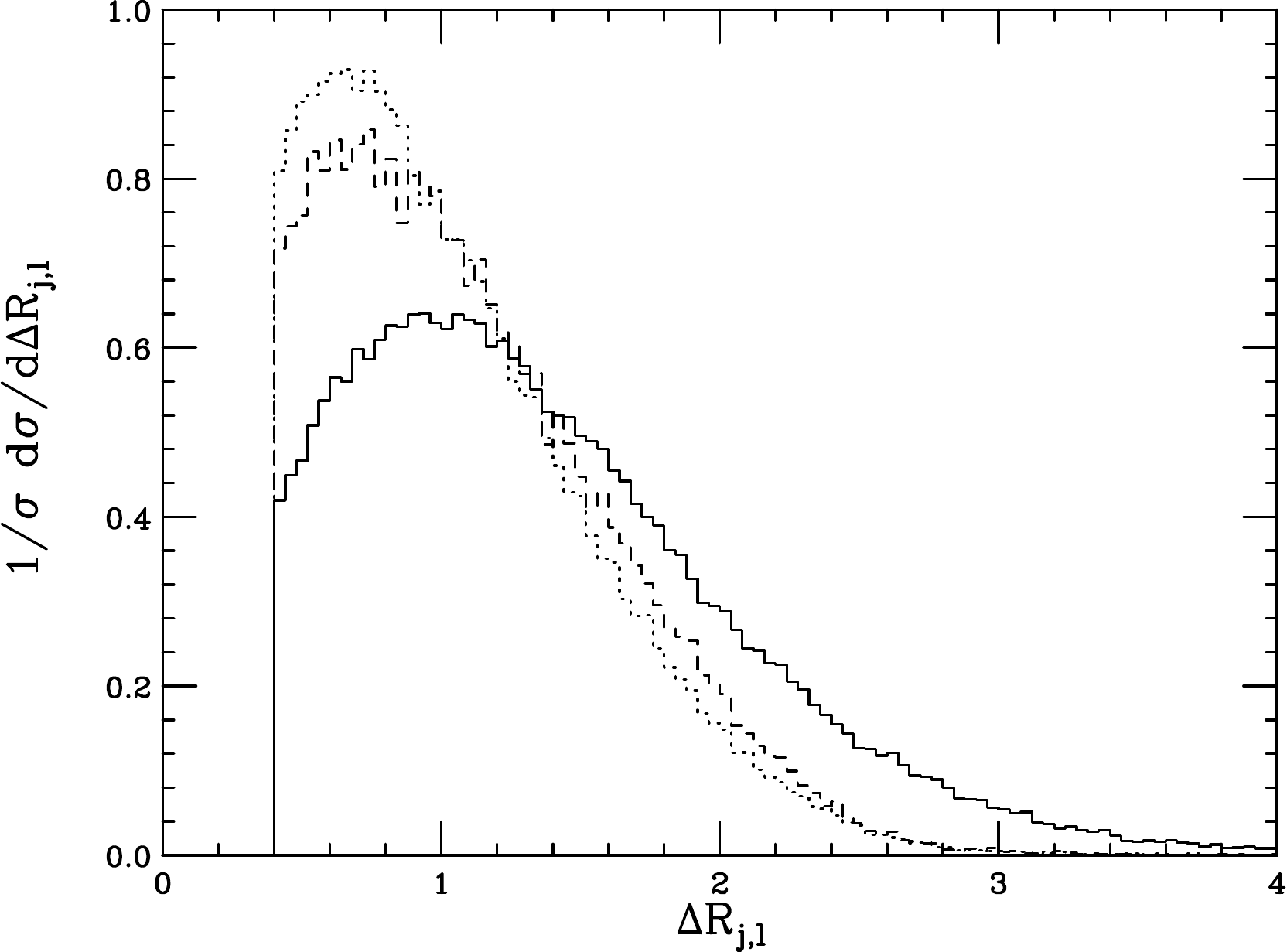}}\hspace{.55cm}
\subfigure[]{\includegraphics[width=0.4\textwidth]{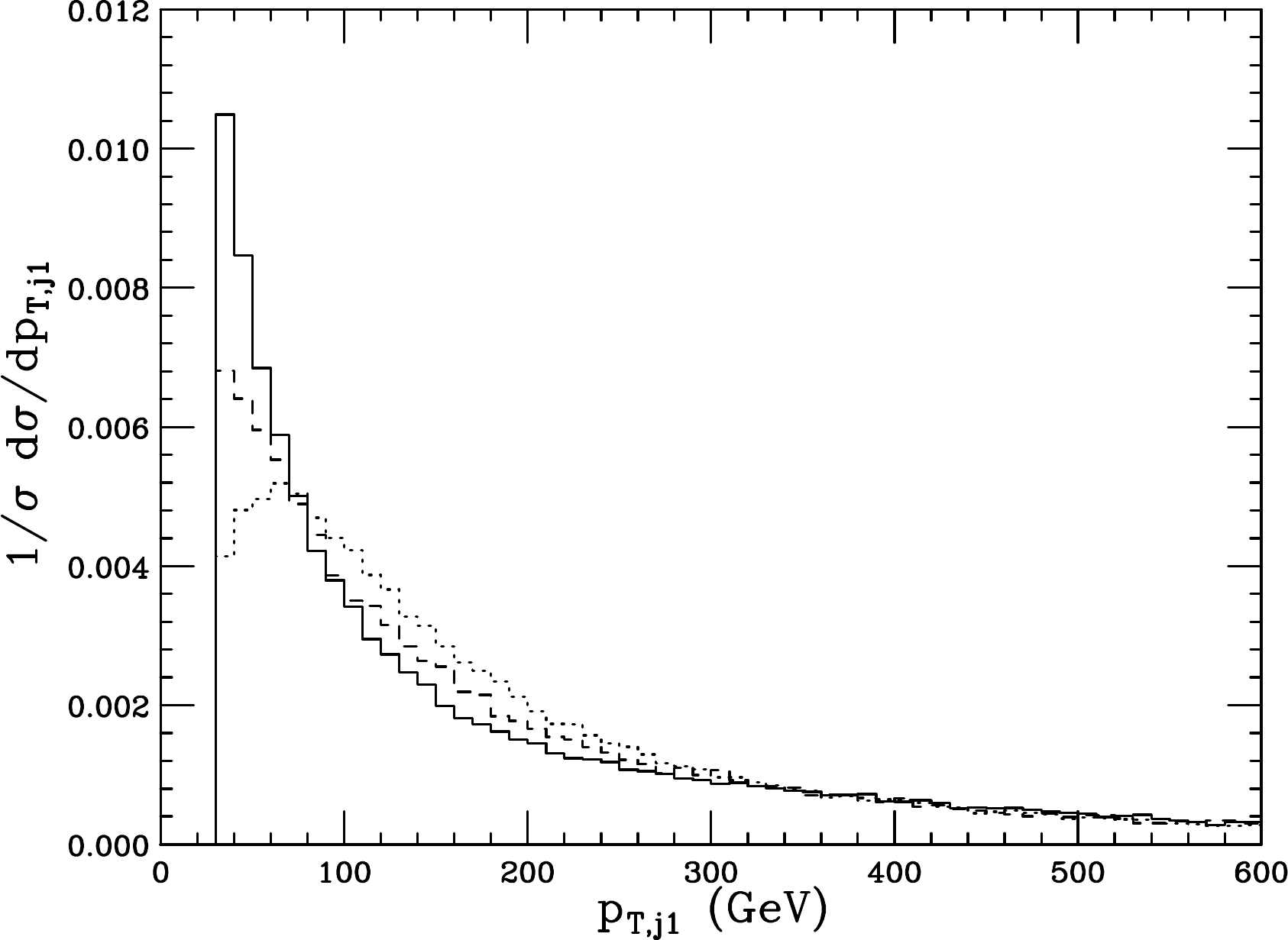}}}
\caption{Results of the simulation:  (a) Lepton transverse momentum
  distribution;
  (b) Lepton pseudorapidity distribution,
  (c) Same-sign lepton-pair
  invariant mass; (d) Angle between same-sign leptons;
  (e) Invariant opening angle $\Delta R_{j\ell}$ between the hardest jet
  and its closest lepton;
  and (f) $p_T$ of the hardest jet $j_1$.
  The solid histograms are the bilepton signals, the dashes correspond to
  the $ZZ$ background, the dots to $t\bar tZ$ processes.}
   \label{results}
\end{figure}
Fig.~\ref{results} shows the results of our simulation for
the  signal (solid histograms), as well as $ZZ$ (dashes) and $t\bar tZ$
(dots) backgrounds.
In detail, we present the spectra of the
lepton transverse momentum $p_T$ (a), pseudorapidity $\eta$ (b),
$\ell^\pm\ell^\pm$ invariant mass $m_{\ell\ell}$ (c),
angle between same-sign leptons $\theta_{\ell\ell}$ (d),
invariant opening angle $\Delta R_{j\ell}$ between the hardest
jets and its closest lepton (e), and hardest-jet transverse momentum
$p_{T,j1}$ (f). We have plotted everywhere normalized
distributions, such as $1/\sigma (d\sigma/dp_T)$, but the normalization of
our spectra to the total cross sections computed above
is quite straightforward. Moreover,
because of the symmetry of our final states, we have included in
the histograms all four leptons and we have suitably normalized the distributions.

A general feature of our results is that our bilepton signal can be
easily separated from the background.
In particular, as for the $p_T$ spectrum, the background distributions
are peaked at low $p_T$ and vanish for $p_T>300$~GeV, while
the rate yielded by the bilepton model is substantial up to about
1.3~TeV.

Our results show that the Bilepton Model
model predicts a higher event fraction with
leptons at central rapidities, say $|\eta|<1$, and relative
angles in the range $1<\theta_{\ell\ell}<2$ respect to other kinematical configurations. The invariant mass $m_{\ell\ell}$
distributions are indeed very different. The signal peaks at
$m_{Y^{++}}\simeq 873$~GeV and manifests as a narrow resonance, whereas the backgrounds
yield broader spectra, peaked around 80 GeV, and roughly vanishing
for $m_{\ell\ell}> 350$~GeV. The $\Delta R_{j\ell}$ and $p_{tj1}$ spectra
are less different than the others, but nevertheless there is a visible discrepancy at low $\Delta R_{j\ell}$ and $p_{tj1}$.

\section{Discussion}

The most popular BSM models, supersymmetry and large
extra dimensions, although not yet excluded,
have received no encouragement from the late
LHC data, since no new particle has so far been discovered.

A more conservative approach is the model of \cite{PHF},
whose phenomenology has been examined in this paper.
Of the new particles of the model, the bilepton gauge bosons
have striking signatures at the LHC. As clear
from Fig.\ref{results}, the large transverse-momentum
rate sets the
clearest distinction of this model respect to the SM.
Specifically, by imposing a cut in $p_{T}$ enforcing a large transverse
momentum above 300 GeV, or even
$p_{T} > 500 $ GeV, the SM background
can be successfully suppressed to the extent that the bilepton
signal, if it is present, cannot be missed in the experimental analysis.

Another hint in this direction, also evident from Fig.\ref{results},
is to focus on the central region with pseudorapidity $|\eta| < 1$
or relative angle $1<\theta_{\ell\ell}<2$, 
which further enhances the rates for this model with respect to the SM ones. The
reason is physically clear when one realizes that the massive bileptons
are produced with rather low momenta in the centre-of mass frame, which
imply central pseudorapidities in the laboratory frame and
larger angles between same-sign leptons. 
Assuming that the four leptons, accompanied by two jets, are muons,
then the invariant
mass of di-muon pairs, with the three different permutational
combinations (12), (13), (14), can be analyzed as in Fig. \ref{results}
to discover decisively the existence of such bilepton particles.
With the magnetic fields available at LHC, namely 4 Tesla at CMS
and 2 Tesla at ATLAS, the stiff muons may still reveal their electric
charges, so that the checking of all three permutations may be obviated.

We therefore believe that searching for bileptons 
is interesting and quite feasible at the LHC. Our investigation
can therefore be seen as a useful starting point to carry out
an experimental analysis on the Bilepton Model \cite{PHF} and its phenomenology, as we have undertaken a full Monte Carlo implementation
of both signal and background, including parton shower and hadronization,
which allows a straightforward application
to the experimental searches.
We hope to return in the future with further investigations which may guide and motivate further analyses.

\section*{Acknowledgements}
We acknowledge discussions with Marco Zaro on the use of the \texttt{MadGraph}
generator and with Gabriele Chiodini, Stefania Spagnolo and Konstantinos Bachas on 
the experimental applications of our analysis.
PHF thanks the INFN and the Department of Physics at the University of Salento
for hospitality. C.C. thanks Rhorry Gauld for discussions.
The work of C.C. is partially supported by INFN `Iniziativa Specifica' QFT-HEP.

\appendix

\section{Vertices}

We give the relevant vertices for the bileptons.

\bea
\ell \;\ell \; Y^{++}=\left\{
\begin{array}{cl}
-\frac{i}{\sqrt2}g_2 \gamma^\mu& P_L\\
\frac{i}{\sqrt2}g_2 \gamma^\mu& P_R
\end{array}
\right.
\eea

\bea
\bar d \;T \; Y^{--}=\left\{
\begin{array}{cl}
-\frac{i}{\sqrt2}g_2 \gamma^\mu &P_L\\
0& P_R
\end{array}
\right.
\eea

\bea
\bar D \;u \; Y^{--}=\left\{
\begin{array}{cl}
\frac{i}{\sqrt2}g_2 \gamma^\mu &P_L\\
0& P_R
\end{array}
\right.
\eea

\bea
h_i\;Y^{++}Y^{--}=\frac{i}{2}g_2^2\left(v_\rho \mathcal{R}^S_{i1}+v_\chi\mathcal{R}^S_{i3}\right)
\eea

\bea
\gamma\;Y^{++}Y^{--}=-2i\,g_2\,\sin\theta_W
\eea

\bea
Z\;Y^{++}Y^{--}=\frac{i}{2}\,g_2\,(1-2\cos2\theta_W)\sec\theta_W
\eea

\bea
Z'\;Y^{++}Y^{--}=-\frac{i}{2}\,g_2\,\sqrt{12-9\sec^2\theta_W}
\eea

\end{document}